\documentclass[iop,numberedappendix]{emulateapj}
\usepackage{graphicx}
\usepackage{rotating}
\usepackage{verbatim}
\usepackage{amsmath}
\usepackage{amssymb}
\usepackage{natbib}
\usepackage[colorlinks=true,
            linkcolor=blue,
            urlcolor=blue,
            citecolor=blue]{hyperref}
\usepackage{mathrsfs}
\usepackage{mathptmx}
\usepackage{array}
\usepackage{wrapfig,lipsum,booktabs}
\usepackage{lineno}

\def\me{m_{\rm e}}

\def\sigmaT{\sigma_{\rm T}}

\def\me{m_{\rm e}}

\def\ba{\begin{align}}
\def\ea{\end{align}}

\def\E{{\cal E}}
\def\mprot{m_{\rm p}}

\def\Rpm{R_{\rm load}}
\def\Rdec{R_{\rm dec}}
\def\tdec{t_{\rm dec}}
\def\Gjet{\Gamma_{\rm jet}}
\def\EGRB{{\cal E}_{\rm GRB}}
\def\Ekin{{\cal E}_{\rm kin}}
\def\Emax{E_{\rm max}}
\def\gth{\gamma_{\rm th}}
\def\epse{\varepsilon_{\rm e}}
\def\mue{\mu_{\rm e}}
\def\tpm{t_{\rm load}}
\def\A{A_{\rm eff}}
\def\epsTeV{\varepsilon_{\rm TeV}}

\newcommand{\OW}{}
\newcommand{\OWS}{}
\newcommand{\OWT}{}
\def\FIC{{\cal F}_{\rm IC}}
\def\Ldiss{L_{\rm diss}}
\def\tauEBL{\tau_{\rm EBL}}
\def\bX{\beta_{\rm X}}

\def\tIC{t_{\rm IC}}
\def\tdyn{t_{\rm dyn}}
\def\uX{u_{\rm X}}
\def\LX{L_{\rm X}}

\slugcomment{2018 ApJ in press}

\shorttitle{A strong limit on the very\OW{-}high-energy emission \OW{from} GRB~150323A}
\shortauthors{The VERITAS Collaboration, Vurm \& Beloborodov}

\begin{document}

\title{A strong limit on the very{-}high-energy emission from GRB~150323A}



\author{
A.~U.~Abeysekara\altaffilmark{1},
A.~Archer\altaffilmark{2},
W.~Benbow\altaffilmark{3},
R.~Bird\altaffilmark{4},
R.~Brose\altaffilmark{5,6},
M.~Buchovecky\altaffilmark{4},
V.~Bugaev\altaffilmark{2},
M.~P.~Connolly\altaffilmark{7},
W.~Cui\altaffilmark{8,9},
M.~Errando\altaffilmark{2},
A.~Falcone\altaffilmark{10},
Q.~Feng\altaffilmark{11},
J.~P.~Finley\altaffilmark{8},
A.~Flinders\altaffilmark{1},
L.~Fortson\altaffilmark{12},
A.~Furniss\altaffilmark{13},
G.~H.~Gillanders\altaffilmark{7},
M.~H\"utten\altaffilmark{6},
D.~Hanna\altaffilmark{11},
O.~Hervet\altaffilmark{14},
J.~Holder\altaffilmark{15},
G.~Hughes\altaffilmark{3},
T.~B.~Humensky\altaffilmark{16},
C.~A.~Johnson\altaffilmark{14},
P.~Kaaret\altaffilmark{17},
P.~Kar\altaffilmark{1},
N.~Kelley-Hoskins\altaffilmark{6},
M.~Kertzman\altaffilmark{18},
D.~Kieda\altaffilmark{1},
M.~Krause\altaffilmark{6},
F.~Krennrich\altaffilmark{19},
M.~J.~Lang\altaffilmark{7},
T.~T.Y.~Lin\altaffilmark{11},
G.~Maier\altaffilmark{6},
S.~McArthur\altaffilmark{8},
P.~Moriarty\altaffilmark{7},
R.~Mukherjee\altaffilmark{20},
S.~O'Brien\altaffilmark{21},
R.~A.~Ong\altaffilmark{4},
N.~Park\altaffilmark{22},
J.~S.~Perkins\altaffilmark{23},
A.~Petrashyk\altaffilmark{16},
M.~Pohl\altaffilmark{5,6},
A.~Popkow\altaffilmark{4},
E.~Pueschel\altaffilmark{6},
J.~Quinn\altaffilmark{21},
K.~Ragan\altaffilmark{11},
P.~T.~Reynolds\altaffilmark{24},
G.~T.~Richards\altaffilmark{25},
E.~Roache\altaffilmark{3},
C.~Rulten\altaffilmark{12},
I.~Sadeh\altaffilmark{6},
M.~Santander\altaffilmark{20},
G.~H.~Sembroski\altaffilmark{8},
K.~Shahinyan\altaffilmark{12},
J.~Tyler\altaffilmark{11},
S.~P.~Wakely\altaffilmark{22},
O.~M.~Weiner\altaffilmark{16},
A.~Weinstein\altaffilmark{19},
R.~M.~Wells\altaffilmark{19},
P.~Wilcox\altaffilmark{17},
A.~Wilhelm\altaffilmark{5,6},
D.~A.~Williams\altaffilmark{14},
B.~Zitzer\altaffilmark{11} (VERITAS Collaboration),\\
and \\
Indrek Vurm\altaffilmark{16}, Andrei Beloborodov\altaffilmark{16}
}

\altaffiltext{1}{Department of Physics and Astronomy, University of Utah, Salt Lake City, UT 84112, USA}
\altaffiltext{2}{Department of Physics, Washington University, St. Louis, MO 63130, USA}
\altaffiltext{3}{Fred Lawrence Whipple Observatory, Harvard-Smithsonian Center for Astrophysics, Amado, AZ 85645, USA}
\altaffiltext{4}{Department of Physics and Astronomy, University of California, Los Angeles, CA 90095, USA}
\altaffiltext{5}{Institute of Physics and Astronomy, University of Potsdam, 14476 Potsdam-Golm, Germany}
\altaffiltext{6}{DESY, Platanenallee 6, 15738 Zeuthen, Germany}
\altaffiltext{7}{School of Physics, National University of Ireland Galway, University Road, Galway, Ireland}
\altaffiltext{8}{Department of Physics and Astronomy, Purdue University, West Lafayette, IN 47907, USA}
\altaffiltext{9}{Department of Physics and Center for Astrophysics, Tsinghua University, Beijing 100084, China.}
\altaffiltext{10}{Department of Astronomy and Astrophysics, 525 Davey Lab, Pennsylvania State University, University Park, PA 16802, USA}
\altaffiltext{11}{Physics Department, McGill University, Montreal, QC H3A 2T8, Canada}
\altaffiltext{12}{School of Physics and Astronomy, University of Minnesota, Minneapolis, MN 55455, USA}
\altaffiltext{13}{Department of Physics, California State University - East Bay, Hayward, CA 94542, USA}
\altaffiltext{14}{Santa Cruz Institute for Particle Physics and Department of Physics, University of California, Santa Cruz, CA 95064, USA}
\altaffiltext{15}{Department of Physics and Astronomy and the Bartol Research Institute, University of Delaware, Newark, DE 19716, USA}
\altaffiltext{16}{Physics Department, Columbia University, New York, NY 10027, USA \email{omw2107@columbia.edu, indrek.vurm@gmail.com}}
\altaffiltext{17}{Department of Physics and Astronomy, University of Iowa, Van Allen Hall, Iowa City, IA 52242, USA}
\altaffiltext{18}{Department of Physics and Astronomy, DePauw University, Greencastle, IN 46135-0037, USA}
\altaffiltext{19}{Department of Physics and Astronomy, Iowa State University, Ames, IA 50011, USA}
\altaffiltext{20}{Department of Physics and Astronomy, Barnard College, Columbia University, NY 10027, USA \mbox{\color{blue}rmukherj@barnard.edu}}
\altaffiltext{21}{School of Physics, University College Dublin, Belfield, Dublin 4, Ireland}
\altaffiltext{22}{Enrico Fermi Institute, University of Chicago, Chicago, IL 60637, USA}
\altaffiltext{23}{N.A.S.A./Goddard Space-Flight Center, Code 661, Greenbelt, MD 20771, USA}
\altaffiltext{24}{Department of Physical Sciences, Cork Institute of Technology, Bishopstown, Cork, Ireland}
\altaffiltext{25}{School of Physics and Center for Relativistic Astrophysics, Georgia Institute of Technology, 837 State Street NW, Atlanta, GA 30332-0430}


\begin{abstract}
On 2015 March 23, VERITAS responded to a {\it Swift}-BAT detection of a gamma-ray burst, with \OW{observations} beginning 270 seconds after the onset of BAT emission, \OW{and only 135 seconds after the main BAT emission peak}. \OW{No statistically significant signal is detected above 140 GeV}.
\OW{The VERITAS upper limit on the fluence in a 40 minute integration corresponds to about 1\% of the prompt fluence}. Our limit is \OW{particularly} significant since the \OW{very-high-energy (VHE)} observation started only $\sim$2 minutes after the prompt emission peaked,
\OW{and} {\it Fermi}-LAT observations of numerous other bursts have revealed
that the high-energy emission is typically delayed relative to the prompt radiation and
lasts significantly longer.
Also, the proximity of GRB~150323A ($z=0.593$) limits the attenuation by the extragalactic background light to $\sim 50$~\% at 100-200 GeV.
We conclude that GRB~150323A \OWT{had an intrinsically very weak high-energy afterglow, or that the GeV
spectrum had a turnover below $\sim100$ GeV.}
If the GRB exploded into the stellar wind of a massive progenitor, the VHE non-detection constrains the wind density parameter to be
$A\gtrsim 3\times 10^{11}$~g~cm$^{-1}$, consistent with a standard Wolf-Rayet progenitor.
Alternatively, the VHE emission from the blast wave would be weak in a very tenuous medium such as the ISM,
which therefore cannot be ruled out as the environment of GRB~150323A.

\end{abstract}


\keywords{gamma-ray burst: general
}

\defcitealias{sheth08}{S08}

\section{Introduction} 
 
\subsection{High-energy radiation from gamma ray bursts}

Gamma-ray bursts (GRB\OW{s}) are \OW{thought to be} powered by ultrarelativistic jets associated with the birth of a compact object.
The bulk of their radiation is typically received over several seconds (the so-called prompt emission),
with spectral peaks clustering around a few hundred keV. 
In contrast, the more long-lived afterglows have been observed across the entire electromagnetic spectrum--from radio to GeV gamma rays.
In particular, the {\it Fermi} Large Area Telescope (LAT) detects approximately $10$ GRBs per year, or roughly 10\% of GRBs that occur in its field of view \citep{Ackermann2013}.
The photon indices measured by LAT cluster around $\Gamma=2$ (i.e. constant energy per logarithmic frequency interval),
without a high-energy spectral break or cutoff; this suggests that substantial energy could be emitted above $\sim 100$~GeV, where it could be detected by
imaging atmospheric Cherenkov telescopes (IACTs). \OW{LAT-detected afterglows roughly decay as 1/t with no clear cutoff and are often observed for hundreds of seconds before the emission becomes too faint for detection. In the case of the bright and nearby GRB 130427A, LAT detected the afterglow for several hours \citep{Ackermann2014}.}

The main advantage of Cherenkov instruments is their large effective area, several orders of magnitude above \OW{space-based} instruments such as LAT,
which more than compensates for the smaller photon flux at very high energies \OW{(VHE, $E > 100$ GeV)} unless the spectrum is extremely steep.
Gamma-ray \OW{burst locations} have indeed been observed by IACTs, and are considered a high-priority target.
However, none has been detected to date \OW{(e.g., \cite{Acciari2011}, \cite{Albert2007}, \cite{Aharonian2009})}. 
Air-shower detectors which are most sensitive at energies above $\sim 10$~TeV have also failed to conclusively detect any of the bursts they observed (e.g. \cite{Abdo2007}, \OWT{\cite{Alfaro2017}}). There was a hint of possible emission from GRB 970417A\OW{;} however the significance of the signal was not considered high enough to indicate unambiguous detection \citep{Atkins2000}. Overall, these non-detections most likely imply a break in the high-energy spectrum in most GRBs. 

The start of IACT observations is typically delayed by a few minutes relative to the prompt trigger, when the GRB has usually already entered the afterglow stage.
In a sparse environment like the interstellar medium (ISM), such delays are comparable to the time it takes for the jet to transfer a sizable fraction of its kinetic energy to the external medium via the forward shock. Furthermore, given the typical jet Lorentz factors of a few hundred, the {\it average} energy available per particle at the shock is
in the TeV range during the early afterglow.
The external blast wave
is thus expected to be a bright TeV emitter
during the first minutes \citep[e.g.][]{Meszaros1994},
{\it regardless of the efficiency} of non-thermal particle acceleration at the shock.
On the other hand, a lack of TeV emission from a bright nearby burst such as GRB~150323A may indicate that the jet has undergone rapid early deceleration in a dense environment such as the stellar wind of the Wolf-Rayet progenitor \citep{Vurm2016}.
Thus, TeV emission constitutes a relatively clean probe of the GRB environment and early blast wave evolution.

\subsection{VERITAS GRB observations}

VERITAS (the Very Energetic Radiation Imaging Telescope Array System) is an IACT array, which is the most sensitive type of instrument for detection of astrophysical gamma-ray emission at
$\sim$1~TeV
energies \citep{Holder2009}. IACT arrays rely on the detection of Cherenkov light \OW{induced} by particles in extensive air showers that were initiated by energetic astrophysical particles entering the atmosphere. The showers are imaged with multiple telescopes \OW{allowing} their incoming directions and energies \OW{to be} reconstructed. VERITAS is sensitive to gamma rays with energies from about 85~GeV to more than 30~TeV \citep{Park2015}. 

When VERITAS receives a burst alert through the GRB Coordinates Network (GCN) \citep{Barthelmy2008}, the on-site observers are prompted to slew the telescopes to the burst position barring any constraints, such as the position of the Moon or the elevation of the burst. The delay between trigger and observation,
which involves the arrival time of
the alert, response by VERITAS observers and telescope slewing, is usually on the order of a few minutes \citep{Acciari2011}. 

Gamma-ray bursts have not been detected at energies greater than $100$~GeV by any instrument to date. Previous observations of {\it Swift} GRBs by VERITAS placed limits on the possibility of particularly strong VHE emission from these bursts \citep{Acciari2011}. At the time of this work, VERITAS has observed more than 150 gamma-ray burst positions, with 50 observations \OW{made} within 180~s of the satellite trigger time. \OWS{Of these,} follow-up observations exist for about 90 bursts detected by {\it Swift}-BAT, 90 bursts detected  by {\it Fermi}-GBM, and 10 bursts detected by {\it Fermi}-LAT. 

A study performed by \cite{Weiner2015} attempted to isolate the most promising gamma-ray burst observations made by VERITAS.
It only
considered
observations of bursts that have been well localized (compared to the VERITAS point spread function) and for which a redshift has been
measured.
It took into account the most important factors (outside of the burst's \OW{VHE energy output}, around which there can be much uncertainty) that impact the observability of a burst:
redshift, \OW{observing} elevation, and observing delay. Eight bursts were identified by this analysis, and
none
was detected individually or with cumulative statistical tests that searched for a faint signal present in multiple observations. The most promising burst observation based on the metrics was GRB~150323A, which is the focus of this paper.

\OW{The data are analyzed using a standard VERITAS analysis package with a selection of analysis parameters that is tuned to soft spectrum-sources, similar to \cite{Aliu2014}, \OWS{where the VERITAS follow-up observation of GRB 130427A is discussed.} (GRB~130427A is a record setting burst that reached the highest observed $\gamma$-ray fluence \citep{Maselli2014}.) The analysis we use here is similarly optimized for a source with an assumed photon index of approximately $3.5$.}
\section{GRB~150323A}

\subsection{Observations of the burst}

\begin{figure}[b]
	\centering
	\centerline{\includegraphics[scale=0.3]{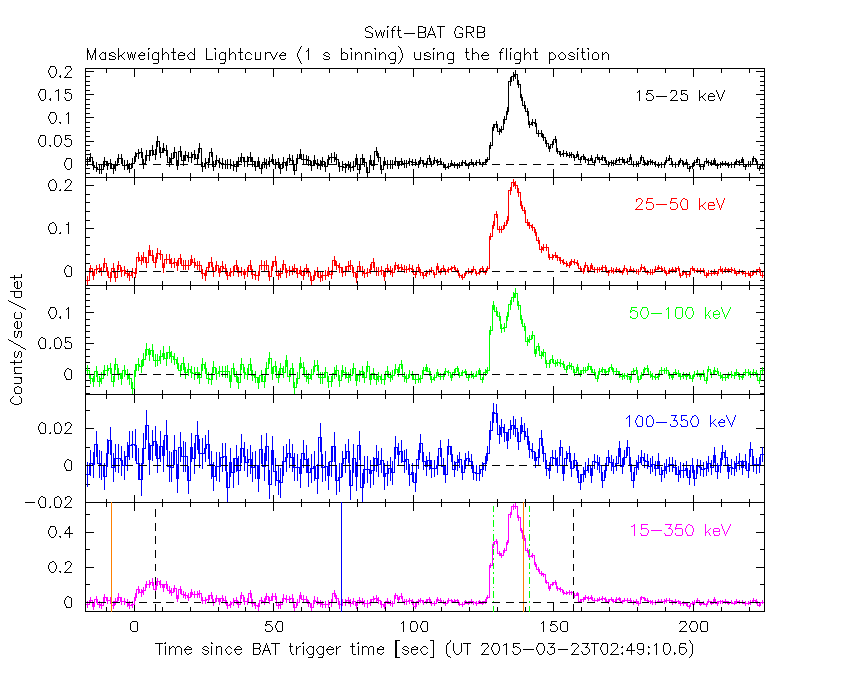}}
	\caption{The {\it Swift}-BAT lightcurve for GRB~150323A, showing both the precursor and the main emission period. The different coloured plots correspond to various energy bands observed by BAT as indicated in each subplot. Taken from the batgrbproduct analysis page: http://gcn.gsfc.nasa.gov/notices\_s/635887/BA/.}
\end{figure}

On 2015 March 23, 02:49:14 UT, the {\it Swift} Burst Alert Telescope (BAT) triggered on a burst with a J2000 position of \OW{(08h 32m 45s.84, 45$^{\circ}$ 26m 02.4s)} and an error radius of approximately 3 arcminutes \citep{Amaral-Rogers2015}. This error radius is both smaller than the VERITAS gamma-ray point spread function ($\sim$0.1 degrees) and the VERITAS field of view ($\sim$3.5 degrees). This position was later refined with {\it Swift} X-ray telescope (XRT) measurements to an accuracy of a few arcseconds \OW{\citep{Goad2015}}. The optical afterglow was detected by the Low Resolution Imaging Spectrometer (LRIS) on the Keck I 10m telescope. Several absorption and emission lines uniformly indicated the redshift of this burst to be $z=0.593$ \citep{Perley2015}. \OWT{A refined analysis by {\it Swift}-BAT \citep{Markwardt2015} found the best fit fluence in the 15-150 keV band to be $6.1 \times 10^{-6}$ erg~cm$^{-2}$. The best-fit photon index in the same spectral window was found to be $1.85$.}

VERITAS began observing the burst 270 seconds after the {\it Swift} trigger. The elevation of the source was 73 degrees at that time, and slowly rising \OWT{(until it reached it's maximum elevation of 76 degrees--an hour later--and began to decline)}. The observations lasted for \OW{170 minutes}. To produce the most sensitive result, we integrated only the first 40 minutes of the observation.\footnote{\OW{The integration time is decided by a Monte Carlo simulation of a reasonable IACT background rate, and a $1/t$ signal at the threshold of detection. The simulation is designed to optimize the \textit{a priori} expected significance for such a signal. One can find the results of an analysis of the full dataset for GRB 150323A in \cite{Weiner2015}.}} This was found to be ideal in the case of a burst that has a flux roughly decaying as $1/t$, as typically found by {\it Fermi}-LAT. \OW{The analysis yielded a result of 71 events in the on-source region, 563 events in the larger region used to estimate the background, and a relative normalization between the two regions, $\alpha$, of 0.132, resulting in a significance of -0.36, using Equation 17 of \cite{Li1983}}. 

We find the VERITAS differential upper limit \OW{($99\%$ confidence level using the method described  in \cite{Rolke2005})} at $140$ GeV is 
\OW{$3.7 \times 10^{-6}$~TeV$^{-1}$~m$^{-2}$~s$^{-1}$}, and the integral upper limit from $140$~GeV to $30$~TeV is \OW{$1.6 \times 10^{-7}$~m$^{-2}$~s$^{-1}$}. This upper limit assumes an intrinsic photon index of $2$, and overlays extragalactic background light (EBL) absorption based on the model described in \cite{Finke2010}\OWS{; attenuation by the EBL increases rapidly above $\sim 100$~GeV and thus softens the observed spectrum of a distant source.}
Alternatively, the $99\%$ confidence level upper limit can be given as
\OW{$19.8$} photons during the first $40$ minutes of VERITAS observation.

\OWT{From the differential upper limit for GRB 150323A at 140~GeV we can calculate a fluence per decade energy by assuming a photon index of 2, giving $6.4\times 10^{-8}$~erg~cm$^{-2}$. This corresponds to about 1 percent of the {\it Swift}-BAT detected prompt fluence.}

VHE photons are known to interact with the EBL to produce electron-positron pairs, attenuating the intrinsic flux appreciably, and making sources difficult to detect from cosmological distances. The resulting gamma-ray attenuation for the VERITAS energy range becomes
large at $z\gtrsim 1$,
although this is somewhat EBL-model dependent.\footnote{\OW{As an example, according to the model described in \cite{Finke2010}, at $z=1$, about $85\%$ of $140$ GeV gamma-rays are absorbed.}}  A redshift of 0.593 is among the lowest typically observed for a GRB \citep{Coward2013}. 

The {\it Swift}-BAT light curve, seen in Figure 1, places GRB~150323A into the ``precursor'' category, where most of the emission is produced tens to hundreds of seconds after a weak trigger event. These types of bursts can account for as few as $3\%$ to as many as $20\%$ of all bursts depending on the criteria used to define them \citep{Burlon2008}.
The light curve of GRB~150323A
consists of one minor peak which triggered the observation, and a larger secondary peak about 135 seconds after the trigger. The VERITAS telescopes were on target 270 seconds after the BAT trigger at 02:53:44 UT, which corresponds to a 135 second delay compared to the main BAT peak. While the VERITAS observation
is delayed relative to the prompt (BAT) emission,
we stress that GeV observations by LAT consistently indicate
a more temporally extended emission at higher photon energies
\citep{Ackermann2013}.
If this result extends to the VERITAS energy band, one would expect strong VHE emission, detectable by VERITAS at the time of observing. \OW{We} note that GRB~150323A \OW{first entered the LAT field of view about an hour after} the {\it Swift}-BAT trigger.

The VERITAS non-detection can be used to explore possible implications for GRB
properties.
We begin with an empirically driven calculation of the expected fluence in the VERITAS energy range,
and conclude that the upper limit is strong and requires a more detailed theoretical analysis.
Then, we discuss
how the expected TeV emission depends on the blast wave energy and GRB environment,
and how the measured upper limit constrains 
the prompt radiative efficiency and the density of the ambient medium.

\subsection{Radiative efficiency in the TeV band}

There are large variations among different bursts in the GeV fluence detected by LAT in comparison to the prompt fluence detected by GBM, and dimmer bursts also have light curves that decay more slowly \citep{Lange2013}. For brighter LAT-detected bursts, the energy emitted in the GeV band clusters around 10\% of the GBM fluence \citep{Ackermann2013}.
Assuming that comparable energy is emitted at higher frequencies,
we calculated the expected fluence in the VERITAS band 
and divided it by the experimental upper limit.
We have assumed that: \\
1) VHE emission begins suddenly and decays as $1/t$.\footnote{We have the emission suddenly end after 1 day, \OW{which is} consistent with the typical \OW{duration} of a LAT observation.} \\
2) The fluence emitted in the VERITAS energy band is given by 10\% of the prompt fluence detected by the BAT.\footnote{BAT and GBM cover nearby energy bands \citep{Sakamoto2008}, where in fact GBM covers a wider range of energies. In cases where the prompt emission peaks in the non-overlapping GBM band, this assumption is in fact conservative. \OWT{This appears to be the case for GRB 150323A, given the hard 1.85 photon index observed by Swift BAT \citep{Markwardt2015}.}}  \\
3) EBL absorption follows the model by Finke \citep{Finke2010}.
\footnote{\label{note1}\OWT{We use \cite{Finke2010} given that it is one of the more conservative (i.e predicts more attenuation) among the recent EBL models. For example, at our threshold energy of 140 GeV (z=0.6) \cite{Finke2010} give an attenuation factor of 0.47, whereas \cite{Gilmore2012} fiducial model and fixed model, \cite{Dominguez2011}, and \cite{Franceschini2008} give factors of 0.42, 0.56, 0.55 and 0.57, respectively.}} \\

4) We approximate the VERITAS effective area as time independent, while in reality it is very slightly changing during the observation. \\

Of these assumptions, we believe (2) is the most dependent on GRB-environment and theory. Assumption (1) has been established by LAT data as a good approximation,\footnote{LAT results show that in the GeV band, afterglow fluence is comparable to prompt fluence, and decays approximately as $1/t$ \citep{Ackermann2013}.} (3) is in fact considered stringent in light of recent results \citep{Abeysekara2015} and assumption number (4) is a very good approximation used for simplification purposes.

The resulting ratios are greater than 1 (see Table 1), indicating that the VHE emission must be weaker than expected by our extrapolation. This suggests a detailed discussion is needed, which we explore in the next section.

\begin{table}[t]
\begin{center}
    \begin{tabular}{ | >{\centering\arraybackslash}p{3cm} | >{\centering\arraybackslash}p{2cm} | >{\centering\arraybackslash}p{2cm} | }
    \hline
     & \OW{VHE} emission begins 1 s after trigger$\rm ^a$ & \OW{VHE} emission begins \OWS{1 s after} main peak ($135$~s) \\ \hline
    Origin time at trigger & \OWS{~1.1} & \OWS{~2.0} \\ \hline
    Origin time at $135$ s & n/a & \OWS{~1.4} \\ \hline 
    \end{tabular}
 \\
    {\scriptsize $\rm ^a$ Consistent with LAT observations of prompt emission delay \\}
\caption{Ratio of the model fluence
to the VERITAS upper-limit under different assumptions. The origin time corresponds to $t=0$ in the $1/t$ time-decay.
As an example, an origin time of 135s could correspond to a burst that was independent of the triggering emission. The emission start time corresponds to the fluence budget of TeV radiation under assumption (2) \OW{[see text]}.}
\end{center}
\label{VERITAS - ICRC 2015}
\end{table}

\section{Theoretical implications}

The interaction between the relativistic GRB jet and the surrounding medium generates luminous high-energy emission.
LAT\OW{-detected} GeV emission
is well explained as radiation from the GRB blast wave loaded with electron-positron pairs \citep{Beloborodov2014}.
The emission is naturally produced
by inverse Compton (IC) scattering of prompt radiation
by {\it thermal} pairs heated at the forward shock.
The model provides good fits to the GeV data and was verified by the detection of the predicted optical counterparts with
a special \OW{(model derived)} light curve \citep{Hascoet2015}.
In most cases the theoretical spectra extend well above $100$~GeV,
where the emission can last from a few minutes up to a day. Below we use this model to interpret the upper limit for GRB~150323A.

Recently, \cite{Vurm2016} 
conducted a systematic study of
both simulated as well as observed GRBs exploding into different media;
they concluded that \OW{the} lack of detections by current Cherenkov instruments
suggests that most of them explode into a dense medium,
such as the stellar wind of the progenitor star.
However, the case of GRB 150323A is somewhat special
owing to its relatively weak X-ray afterglow \citep{Melandri2015},	
resulting in fewer targets for IC emission.
Consequently, one cannot 
conclusively rule out the ISM as the ambient medium in this burst.

\subsection{Wind medium}

Given its relatively modest energy budget $\EGRB \approx 10^{52}$~erg \citep{Golenetskii2015}, 
the jet of GRB 150323A expanding into the dense progenitor wind would have entered
the self-similar deceleration regime by the time the VERITAS observation started.
\OW{By this time,} the dissipated luminosity at the forward shock is approximately
$\Ldiss \sim \Ekin/(4t)$, where $t$ is the time in the cosmological rest frame of the burst.
\OW{Particle-in-cell (PIC)} simulations of collisionless shocks suggest that a fraction $\epse\sim 0.3$ of this energy is placed into heated thermal electrons \citep{Sironi2011}. 
Unless the shock is strongly magnetized, the electrons radiate most of their energy via IC emission.
The IC fluence received over a logarithmic time interval is 
\begin{align}
\FIC \sim \frac{1+z}{4\pi d_L^2}
\, t L_{\rm IC} = 
8.1\times 10^{-7}
\, \E_{{\rm kin},52} \left(\frac{\epse}{0.3} \right) \,\, \mathrm{erg~cm}^{-2},
\end{align}
where $L_{\rm IC}=\epse\Ldiss$ and the normalization of the jet kinetic energy corresponds to 50~\% radiative efficiency ($\E_{{\rm kin}} = \E_{\rm GRB} = 10^{52}$ erg). We use the common notation that $X_{,n}$ corresponds to the quantity $X$ divided by $10^n$ with suitable units so as to make the result dimensionless.
Parametrizing the fraction of the IC energy
that emerges in the VHE band as $\epsTeV$, 
the corresponding photon count at the detector is
\begin{align}
&{\cal N} \sim  \A \frac{\epsTeV\FIC}{\overline{E}_{\rm ph}} \, e^{-\tauEBL}	\nonumber \\
&\approx 85  \,
 \E_{{\rm kin},52}  \left(\frac{\A}{5\times 10^8 \, \mathrm{cm}^2}\right)
\left( \frac{\overline{E}_{\rm ph}}{140\, \mathrm{GeV}} \right)^{-1} \left(\frac{\epse}{0.3} \right) \,
\left( \frac{\epsTeV}{0.1} \right),
\label{eq:count}   
\end{align}
where \OW{$e^{-\tauEBL} \approx 0.47$} accounts for attenuation by
the extragalactic background light at $140$~GeV \citep{Finke2010}.\footnote{\OWT{See footnote 30.}}

The IC spectrum of the thermal electrons has approximately the same slope as the soft target radiation;
the photons upscattered into the VHE band are typically from the X-ray domain.
Given the observed X-ray photon index $\bX\approx 2.0$ \citep{Melandri2015},
the gamma-ray spectrum during the VERITAS observation
is expected to be flat in terms of energy per logarithmic frequency interval.
The spectrum cuts off at
$\Emax = \Gamma\gth\me c^2$, where $\gth$
is the average Lorentz factor of thermal electrons heated in the forward shock.
\OWS{Even if} $\Emax\gg 100$~GeV
during the VERITAS observation,
the observable window is limited\OW{: EBL absorption suppresses emission above $E\sim 300$~GeV,
and the sensitivity of Cherenkov instruments declines below $\sim 100$~GeV.}
In this case $\epsTeV=0.1$ is a reasonable estimate, and Equation (\ref{eq:count}) predicts \OW{about a hundred detectable} counts,
well above the upper limit of \OW{20} from the 40 minute VERITAS observation.
On the other hand, if $\Emax\lesssim 100$~GeV throughout the observation,
then effectively $\epsTeV=0$ and no VHE emission is expected.\footnote{In our discussion we are neglecting
the additional contribution from a possible nonthermal population of {\it accelerated} leptons.
Their energy budget is expected to be significantly lower, but could also contribute to the VHE emission.}
We consider it likely that this is the reason for the non-detection of GRB 150323A by VERITAS:
the thermal IC emission cuts off below 100~GeV,
while the IC component from nonthermal accelerated electrons is too weak to be detected.

Over most of the afterglow stage
the maximal IC photon energy from thermal electrons is controlled by the
Lorentz factor of the forward shock.
However, during the first minute after the explosion the prompt radiation ahead of the forward shock
loads the ambient medium with a large number of pairs (\cite{Thompson2000}; \cite{Meszaros1994}; \cite{Beloborodov2002});
consequently, the average energy {\it per lepton} is low and $\Emax$ is below the VHE band.
The pair loading ends at
(\cite{Thompson2000}; \cite{Beloborodov2002})
\begin{align}
\Rpm = 5.7 \times 10^{15} \, \E_{{\rm GRB}, 52}^{1/2} \,\, \mathrm{cm}.
\label{eq:Rpm}
\end{align}
The turnover of the IC spectrum attains its maximal value at this radius,
\begin{align}
\Emax(\Rpm) &\approx
\left. \Gamma\gth \me c^2 \right|_{\Rpm}
\approx
\left.\Gamma^2 \, \mue \epse \mprot c^2 \right|_{\Rpm}	\nonumber \\
&=440
\, \frac{\E_{{\rm kin},52}}{A_{11} \E_{{\rm GRB}, 52}^{1/2}} \, \left(\frac{\epse}{0.3} \right) \,\, \mathrm{GeV},
\label{eq:Emaxmax}
\end{align}    
where $\gth=\Gamma\mue\epse\mprot/\me$, A is the wind density parameter (a standard density Wolf-Rayet wind has $A\sim 3\times 10^{11}$~g~cm$^{-1}$), $\mue=2$ is the mean molecular weight per proton in a Wolf-Rayet progenitor wind,
and we have used
$\Gamma = [\Ekin/(8\pi c^2 A R)]^{1/2}$ for the self-similarly decelerating blast wave.
This occurs at observer time
\begin{align}
\tpm \approx \left. \frac{(1+z) R}{2c\Gamma^2}\right|_{\Rpm} = 
190 \,
\frac{A_{11}}{\E_{{\rm kin},52}} \, \E_{{\rm GRB},52}	\,\, \mathrm{s}.
\label{eq:tpm}
\end{align}

The VHE IC emission is suppressed at all times if
$\Emax(\Rpm)<(1+z) \times 100$~GeV,
which yields a constraint
\begin{align} 
\frac{\E_{{\rm kin},52}}{A_{11}} \lesssim 0.36 \,\,\, \E_{{\rm GRB},52}^{1/2} \left(\frac{\epse}{0.3} \right)^{-1} \,\, \mathrm{erg}.
\end{align}
This condition is marginally satisfied in a standard density Wolf-Rayet wind with $A\sim 3\times 10^{11}$~g~cm$^{-1}$
if the jet is at least moderately radiatively efficient in the prompt phase,
i.e. $\Ekin \lesssim \EGRB \approx 10^{52}$~erg.

In typical bursts, the pair-production opacity due to the X-ray afterglow photons
suppresses the VHE emission at early times.
However, for GRB 150323A it can be shown that attenuation by intrinsic $\gamma\gamma$-absorption
was at most marginal at $t\gtrsim 300$~s (after the steep early decline of the X-ray lightcurve),
owing to its comparatively weak X-ray afterglow. It can also be shown that in a wind medium
the weak X-ray afterglow nevertheless provides sufficient targets for marginally efficient IC cooling of the VHE emitting electrons.

\subsection{ISM}

In the low density ISM the jet decelerates significantly later than in the wind medium,
and $\Emax\gg 100$~GeV at $\Rpm$.
The dissipation rate at the forward shock peaks at the deceleration radius
\begin{align}
\Rdec = \left(\frac{3\Ekin}{8\pi\mprot c^2 n \Gjet^2}
\right)^{1/3}
=4.0\times 10^{16} \, \frac{\E_{{\rm kin},52}^{1/3}}{n^{1/3} \, \Gamma_{{\rm jet},2}^{2/3}} \,\, \mbox{cm},
\end{align}
where $\Gjet$ is the initial jet Lorentz factor.
The corresponding observer time for redshift $z=0.6$ is $\tdec = 230 \E_{{\rm kin},52}^{1/3}\Gamma_{{\rm jet},2}^{-8/3}n^{-1/3}$~s.
Since $\Rdec > \Rpm$, the VHE emission is also expected to peak near $\Rdec$.

At $R>\Rdec$ the shock-dissipated luminosity is $\Ldiss \sim 3\Ekin/(8t)$,
i.e. comparable to that in the wind medium.
However, owing to the larger characteristic $R$ and $\Gamma$ in the ISM,
along with the weak X-ray afterglow of GRB 150323A,
the shock-heated electrons are unable to cool/radiate efficiently.
The ratio of expansion and IC cooling times for post-shock thermal electrons at $\tdec$ is \citep{Vurm2016}
\begin{align}
\frac{\tdyn}{\tIC} = \frac{4\sigmaT\uX\gth}{3\me c}
= 0.028 \, \frac{n^{1/3}}{\E_{{\rm kin},52}^{1/3} \Gamma_{{\rm jet},2}^{4/3}} \, L_{{\rm X},46} \, \left(\frac{\epse}{0.3} \right).
\label{eq:cool}
\end{align}
Here $\LX$ is X-ray afterglow luminosity that provides the targets for IC scattering
and $\uX = \LX/(4\pi c R^2 \Gamma^2)$ is the comoving radiation energy density;
in Equation (\ref{eq:cool}) $\LX$
is normalized to the observed value
just after the steep decline which ends at $\sim 500$~s.

In the slow-cooling regime
the electrons radiate only a fraction $\sim\tdyn/\tIC$ of their energy before cooling adiabatically,
which amounts to a few percent using our fiducial parameters. 
Including this factor,
the count estimate (\ref{eq:count})
becomes consistent with a non-detection.
Note, however, that the VERITAS observation started during the steep X-ray decay;
the X-ray luminosity was above $10^{47}$~erg~s$^{-1}$ for the first $50$~s of observation.
Although it suggests that VHE gamma rays from this brief epoch could have been detectable,
it does not constitute sufficient evidence
to rule out the ISM as the environment of GRB 150323A.

\section*{Conclusions}

We report the VERITAS observation of GRB 150323A,
a promising candidate for the detection of VHE gamma rays
owing to its relative proximity ($z=0.593$), high \OW{observing} elevation and the rapid response time of VERITAS, $270$~s from the {\it Swift}/BAT trigger.
No statistically significant signal was detected.
We place a $99\%$ confidence level differential upper limit on the 140~GeV fluence at \OW{$6.4\times 10^{-8}$~erg~cm$^{-2}$},
which constitutes \OW{$\sim 1$~\%} of the prompt fluence.
For comparison,
the average GeV fluence of LAT-detected GRBs is $\sim 10$~\% of the prompt \citep{Ackermann2013} (unfortunately, no LAT observations are available for GRB 150323A).
A naive 
extrapolation of the approximately flat spectra typically observed in the LAT band (in terms of $\nu F_{\nu}$)
would place a comparable amount of energy in the VHE band.
The LAT emission usually peaks within the first $\sim 10$~seconds, and decays as $t^{-\alpha}$, where $\alpha\sim 1$.
Thus even accounting for the additional delay, the deep VHE limit for GRB 150323A suggests that either
(1) it had an intrinsically very weak high-energy afterglow (possibly hinted by its weak X-ray afterglow),
or (2) the GeV spectrum had a turnover below $\sim 100$~GeV.

From a theoretical perspective,
the energy dissipated by the relativistic blast wave of GRB 150323A would have been sufficient for a VERITAS detection
if even $\sim 1$~\% of the dissipated energy was radiated in the VHE band,
unless the prompt radiative efficiency was extremely high
(i.e. almost no energy was left for the blast wave, $\Ekin\ll\EGRB$).
Using a ``minimal" model
where the high-energy emission is produced by shock-heated {\it thermal} electrons
upscattering the (observed) X-ray afterglow radiation,
we were able to
place a constraint on the ratio of the blast wave kinetic energy and the ambient medium density.
The high-energy turnover
of the IC spectrum
remains below $100$~GeV throughout the afterglow stage if
GRB 150323A was
a moderately radiatively efficient burst ($\Ekin\approx\EGRB$) exploding into a standard Wolf-Rayet progenitor wind ($A\approx 3\times 10^{11}$~g~cm$^{-1}$).
Alternatively, the blast wave would be dim in the VHE band at the time of the VERITAS observation if
it exploded into a low-density ISM, due to inefficient cooling of the shock-heated electrons.

\section*{Acknowledgments}
VERITAS is supported by grants from the U.S. Department of Energy Office of Science, the U.S. National Science Foundation and the Smithsonian Institution, and by NSERC in Canada. We acknowledge the excellent work of the technical support staff at the Fred Lawrence Whipple Observatory and at the collaborating institutions in the construction and operation of the instrument. The VERITAS Collaboration is grateful to Trevor Weekes for his seminal contributions and leadership in the field of VHE gamma-ray astrophysics, which made this study possible. \OWS{Indrek Vurm acknowledges support from the Estonian Research Council grant PUT1112.}

\end{document}